\documentclass[aps,prl,reprint,groupedaddress,showpacs,preprintnumbers,amsmath,amssymb,superscriptaddress]{revtex4-1}

\usepackage{graphicx}
\usepackage{graphics}
\usepackage{dcolumn}
\usepackage{bm}
\usepackage{sidecap}
\usepackage{booktabs}
\usepackage{tabularx}
\usepackage{epstopdf}
\begin{document}

\title{First Results from the CARIBU Facility: Mass Measurements on the $r$-Process Path}

\author{J. \surname{Van Schelt}}
 \affiliation{Physics Division, Argonne National Laboratory, Argonne, Illinois 60439, USA}
 \affiliation{Department of Physics, University of Chicago, Chicago, Illinois 60637, USA}
\author{D. Lascar}
 \affiliation{Department of Physics and Astronomy, Northwestern University, Evanston, Illinois 60208, USA}
 \affiliation{Physics Division, Argonne National Laboratory, Argonne, Illinois 60439, USA}
\author{G. Savard}
 \affiliation{Physics Division, Argonne National Laboratory, Argonne, Illinois 60439, USA}
 \affiliation{Department of Physics, University of Chicago, Chicago, Illinois 60637, USA}
\author{J. A. Clark}
 \affiliation{Physics Division, Argonne National Laboratory, Argonne, Illinois 60439, USA}
\author{P. F. Bertone}
 \affiliation{Physics Division, Argonne National Laboratory, Argonne, Illinois 60439, USA}
\author{S. Caldwell}
 \affiliation{Department of Physics, University of Chicago, Chicago, Illinois 60637, USA}
 \affiliation{Physics Division, Argonne National Laboratory, Argonne, Illinois 60439, USA}
\author{A. Chaudhuri}
 \affiliation{Department of Physics and Astronomy, University of Manitoba, Winnipeg, Manitoba R3T 2N2, Canada}
 \affiliation{Physics Division, Argonne National Laboratory, Argonne, Illinois 60439, USA}
 \altaffiliation[Now at]{TRIUMF, Vancouver, BC V6T 2A3, Canada}
\author{A. F. Levand}
 \affiliation{Physics Division, Argonne National Laboratory, Argonne, Illinois 60439, USA}
\author{G. Li}
 \affiliation{Department of Physics, McGill University, Montr\'{e}al, Qu\'{e}bec H3A 2T8, Canada}
 \affiliation{Physics Division, Argonne National Laboratory, Argonne, Illinois 60439, USA}
\author{G. E. Morgan}
 \affiliation{Department of Physics and Astronomy, University of Manitoba, Winnipeg, Manitoba R3T 2N2, Canada}
\author{R. Orford}
 \affiliation{Department of Physics, McGill University, Montr\'{e}al, Qu\'{e}bec H3A 2T8, Canada}
\author{R. E. Segel}
 \affiliation{Department of Physics and Astronomy, Northwestern University, Evanston, Illinois 60208, USA}
 \affiliation{Physics Division, Argonne National Laboratory, Argonne, Illinois 60439, USA}
\author{K. S. Sharma}
 \affiliation{Department of Physics and Astronomy, University of Manitoba, Winnipeg, Manitoba R3T 2N2, Canada}
\author{M. G. Sternberg}
 \affiliation{Department of Physics, University of Chicago, Chicago, Illinois 60637, USA}
 \affiliation{Physics Division, Argonne National Laboratory, Argonne, Illinois 60439, USA}

\date{\today}

\begin{abstract}
The Canadian Penning Trap mass spectrometer has made mass measurements of 33 neutron-rich nuclides provided by the new Californium Rare Isotope Breeder Upgrade (CARIBU) facility at Argonne National Laboratory.  The studied region includes the $^{132}$Sn double shell closure and ranges in $Z$ from In to Cs, with Sn isotopes measured out to $A=135$, and the typical measurement precision is at the 100~ppb level or better.  The region encompasses a possible major waiting point of the astrophysical $r$ process, and the impact of the masses on the $r$ process is shown through a series of simulations.  These first-ever simulations with direct mass information on this waiting point show significant increases in waiting time at Sn and Sb in comparison with commonly used mass models, demonstrating the inadequacy of existing models for accurate $r$-process calculations.
\end{abstract}

\maketitle

\emph{Introduction.}---Neutron-rich radioactive nuclides have remained stubbornly inaccessible to precision study for decades, because while fission of heavy elements abundantly produces these nuclides, their chemistry prohibitively impedes the release of all but the most volatile elements from traditional bulky fission sources~\cite{REX-ISOLDE}.  The problem of chemistry dependence in neutron-rich beam production has been abated by the development of gas cells which can stop and thermalize reaction products in gas~\cite{Aysto_IGISOL}, with the most efficient being the new radiofrequency (RF) gas catchers~\cite{Wada_gas_catchers,sa03nimb}.  These developments have allowed comprehensive surveys of the properties of fission products to begin.

A key motivation for the study of these nuclides is the astrophysical rapid neutron-capture process ($r$ process), which is thought to have produced half of the heavy nuclei in the universe~\cite{BBFH,Cowan-rev,Qian_2003,Arnould_rev}.  The site of the $r$ process which populates the universe is not known, but is possibly within core-collapse supernovae or events which eject material from the crusts of neutron stars such as binary mergers.  During an $r$-process event, the balance between neutron capture $(n,\gamma)$ and photodissociation $(\gamma,n)$ reactions determines the distribution of isotope populations within each element, while $\beta$ decay moves nuclei to higher proton numbers.  A critical nuclear physics input to the $(\gamma,n)$ rate is the neutron separation energy ($S_n$) of the participant nucleus, and is calculated from the atomic masses of parent and daughter isotopes.  Direct mass measurements made with Penning traps~\cite{Geonium} are only now reaching nuclides on possible $r$-process paths.  The Canadian Penning Trap mass spectrometer (CPT)~\cite{fa11prc,jv12phd, dl12phd} and other Penning traps have made precision mass measurements of progressively more neutron-rich nuclides~\cite{fission_mass_review,sa06ijmsip,jv12prc} as the state of the art of gas-catcher technology has advanced.  

\begin{figure}[bht]
\centering
\includegraphics[width=0.5\textwidth]{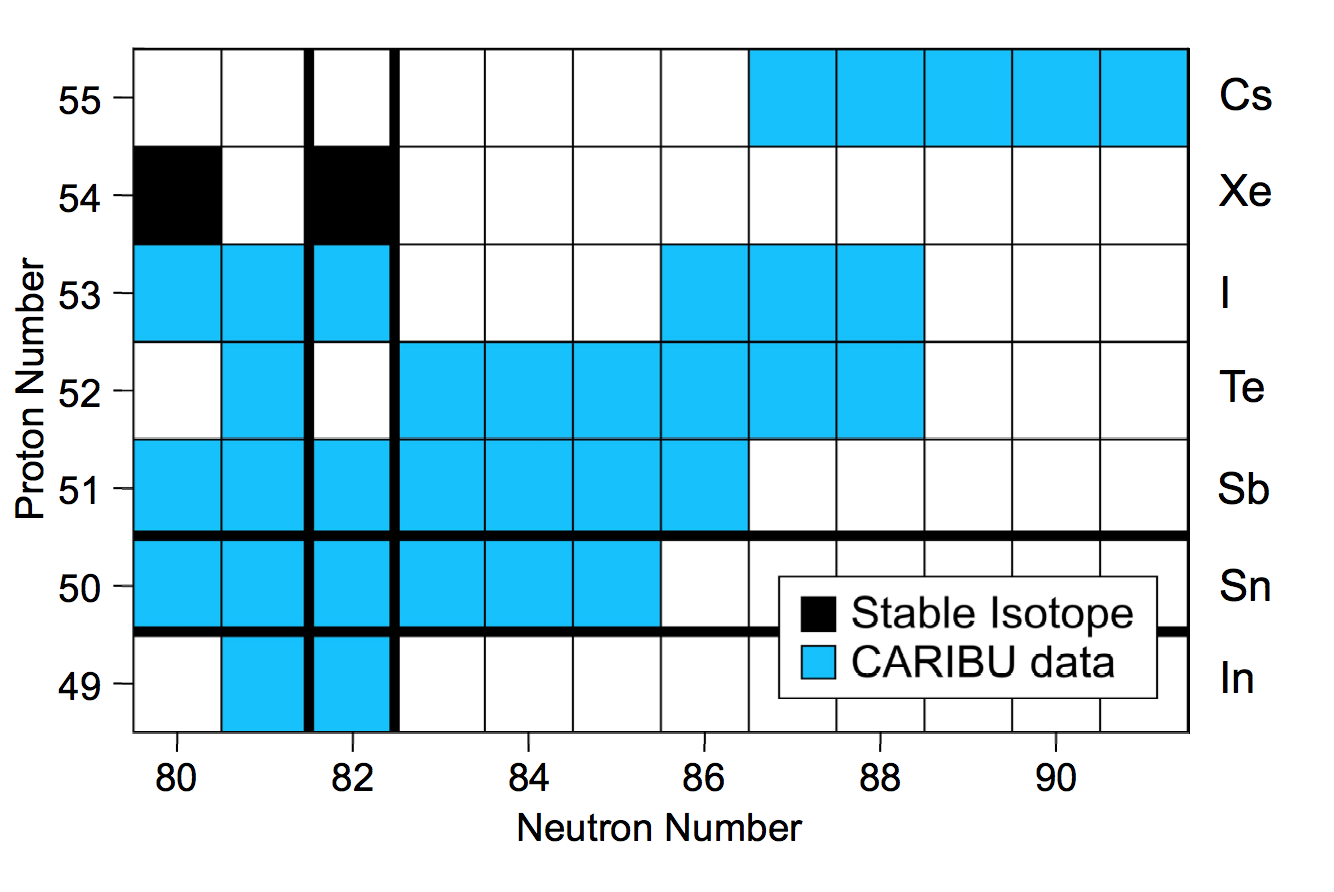}
\caption{\label{fig:chart}(Color online) Locations of the nuclides for which masses have been measured by the CPT at CARIBU.  The shell closures at $N=82$ and $Z=50$ are highlighted.}
\end{figure}

RF gas catchers are now sufficiently mature to justify new facilities dedicated to their beams. The new Californium Rare Isotope Breeder Upgrade (CARIBU)~\cite{CARIBU} is the first such facility, now online.  CARIBU uses a $^{252}$Cf spontaneous fission source in a large gas catcher to provide purified neutron-rich beams to either low-energy beam experiments or the Argonne Tandem-Linac Accelerator System (ATLAS) for reaccelerated beam experiments.  This paper describes the first science results using beams from CARIBU: mass measurements made with the CPT of the 33 nuclides shown in Fig.~\ref{fig:chart}, 12 of which improve the precision over earlier attempts.

\emph{The Measurements.}---For the bulk of these measurements CARIBU used a 50~mCi fission source, providing beams to the CPT as intense as 4200 ions/s of $^{142}$Cs, cleaned by the isobar separator~\cite{CARIBU_IS} with a resolution of $\delta m/m\approx 1/9000$.

The CPT was moved to the CARIBU low-energy experimental area and recommissioned in 2011.  Upgrades made during the move include the installation of a $^{133}$Cs$^+$ ion source for calibration and tuning, and liquid nitrogen cooling of the preparation linear RFQ ion trap located just before the CPT.  Ions are ejected from the CARIBU buncher every 100--200~ms, and are accumulated and cooled in the preparation trap until transfer to the CPT.

\begin{figure}
\centering
\includegraphics[width=0.5\textwidth]{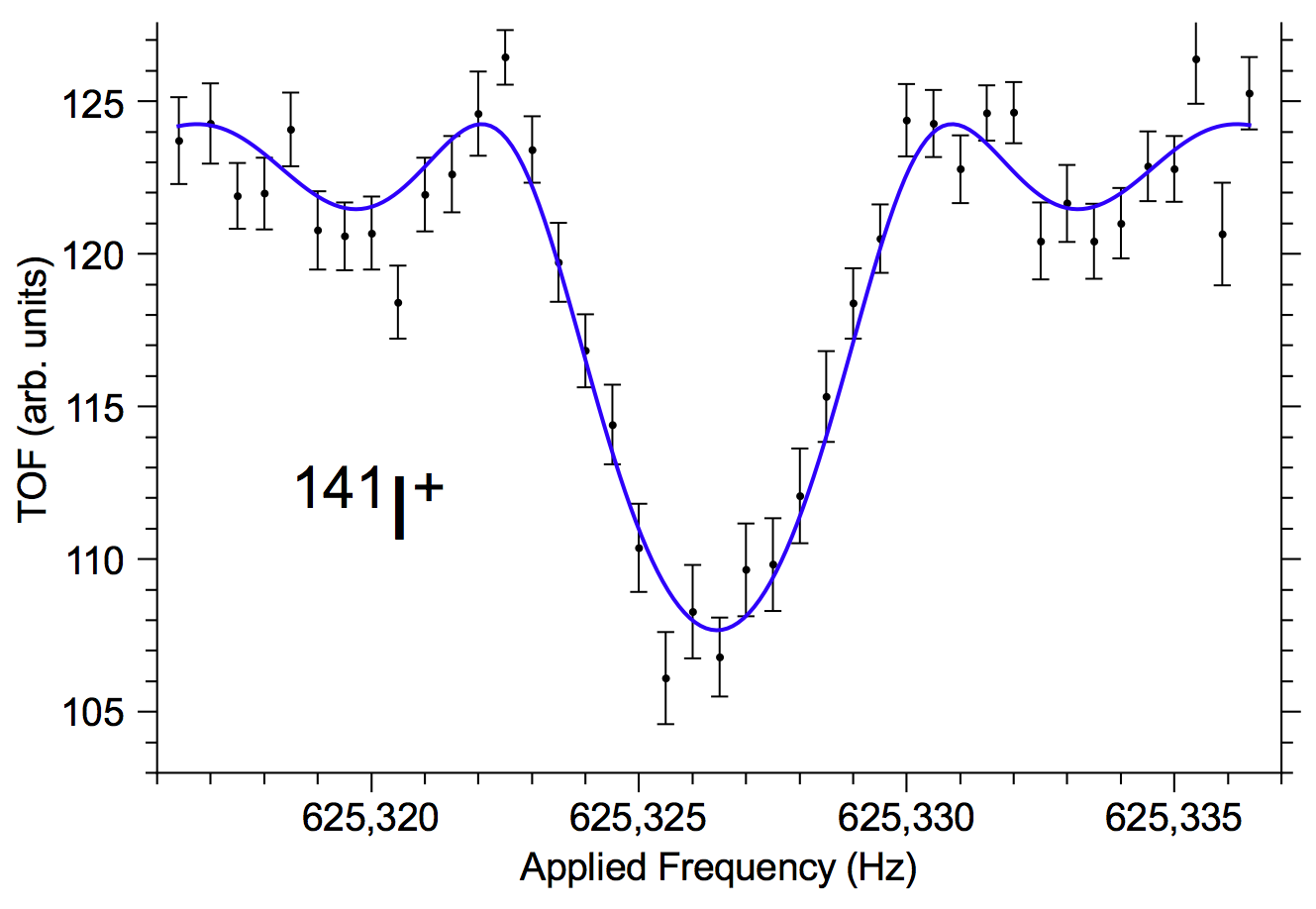}
\caption{\label{fig:141I}(Color online) Example data, a \hbox{200-ms} excitation of $^{141}$I$^+$.  The points represent the time-of-flight data, and the curve a fit to the theoretical response.}
\end{figure}

The CPT uses the time-of-flight ion cyclotron resonance method of Penning trap mass spectrometry for its measurements~\cite{Graff, Bollen_accuracy, Konig_conversion}.  The cyclotron frequency \mbox{$\omega_c=qB/m$} (where $q$ is charge, $B$ magnetic field strength, $m$ mass) is measured through successive attempts to resonantly convert ion bunches from slow to fast orbital motion.  Ions are ejected out of the magnet to a detector, producing a time-of-flight versus applied frequency spectrum as shown in Fig.~\ref{fig:141I}.  The spectrum is fit with an approximation of the theoretical response in~\cite{George_damping}, with the minimum time of flight occurring at the cyclotron frequency. The mass is then determined from the ratio of the measured cyclotron frequency of a well-known calibrant species to that of the species of interest.

A strength of Penning trap measurements is their small systematic errors~\cite{Bollen_accuracy}, which were investigated again for the recommissioning of the CPT.  The superconducting magnet's field strength exhibited more scatter over time than before its move to CARIBU, such that the 11~ppb standard deviation among the calibrations was added as a systematic uncertainty to all measurements.  Frequency shifts linear in mass difference are possible due to trap imperfections~\cite{Geonium}, and a search found such a shift in the CPT, compensated for in the analysis by adjusting frequencies by $\frac{\Delta f}{f}=(A-133)\times1.74(18)$~ppb, where $A$ is the mass number of the unknown.  Finally, the detected ion rate was targeted below 6 detected ions per bunch in the trap to prevent ion-ion interactions~\cite{Bollen_isomer} from introducing significant systematic effects.

\begin{table*}[tbh] 
\centering 
\caption{\label{tbl:masses}CPT results from CARIBU beams. Mass values are combined with earlier CPT work where indicated, while all cyclotron frequency ratios are strictly from the new CARIBU-based results. Differences are shown from the 2003 Atomic Mass Evaluation (AME)~\cite{AME03_1,*AME03_2} and other direct measurements from JYFLTRAP~\cite{JYFLTRAP_GS}, ISOLTRAP~\cite{99Am05,ISOL-Cs08,ISOL-Sn1,ISOL-Sn2}, and the FRS-ESR ring~\cite{GSI_08,GSI_12}. In the difference columns, uncertainties in parentheses are the combined uncertainties, and those in braces are of the sources cited. The 2003 AME is shown rather than the new 2012 AME~\cite{AME12_1,*AME12_2} due to many of the latter's entries being dominated by the measurements shown in columns 5--7 and in~\cite{jv12prc}.}
\begin{tabular*}{\textwidth}{@{\extracolsep{\fill}} l l l r@{\extracolsep{0 pt}}l @{\extracolsep{\fill}} r@{\extracolsep{0 pt}}l @{\extracolsep{\fill}} c r@{\extracolsep{0 pt}}l @{\extracolsep{\fill}} c r@{\extracolsep{0 pt}}l c} 
\hline\hline\noalign{\smallskip}
 & \multicolumn{1}{c}{$r=\frac{\omega_c\left(^{133}\textrm{Cs}^+\right)}{\omega_c(\textrm{Unknown}^+)}$} & \multicolumn{12}{c}{Mass Excess of Neutral Atom (keV)}\\ 
\noalign{\smallskip}\cline{3-14}\noalign{\smallskip}
\multicolumn{1}{l}{Nuclide} & \multicolumn{1}{c}{CPT at CARIBU} & \multicolumn{1}{c}{CPT} & \multicolumn{2}{c}{$\Delta_{\textsc{CPT}-\textsc{AME03}}$} & \multicolumn{2}{c}{$\Delta_{\textsc{CPT}-\textsc{JYFLTRAP}}$} & ref. & \multicolumn{2}{c}{$\Delta_{\textsc{CPT}-\textsc{ISOLTRAP}}$} & ref. & \multicolumn{2}{c}{$\Delta_{\textsc{CPT}-\textsc{FRS-ESR}}$} & ref. \\
\noalign{\smallskip}\hline\noalign{\smallskip}
$^{130}$In$^\textrm{a}$ & 0.977\,576\,25(16) & $ -69\,652(20) $ & $ 238 $ & $ (44) \{40\} $ & $ $ & $  $ & & $ $ & $  $ & & $ $ & $  $ & \\
$^{131}$In$^\textrm{a}$ & 0.977\,575\,78(16) & $ -67\,876(35) $ & $ 262 $ & $ (45) \{28\} $ & $ 149 $ & $ (35) \{2.6\} $ & \cite{JYFLTRAP_GS} & $ $ & $  $ & & $ $ & $  $ & \\
\noalign{\smallskip}                          
$^{130}$Sn$^\textrm{b}$ & 0.977\,491\,606(29) & $ -80\,130.8(3.6) $ & $ 8 $ & $ (11) \{11\} $ & $ 2 $ & $ (5) \{4\} $ & \cite{JYFLTRAP_GS} & $ -3 $ & $ (16) \{16\} $ & \cite{ISOL-Sn1} & $ $ & $  $ & \\
$^{131}$Sn$^\textrm{b}$ & 0.985\,038\,975(35) & $ -77\,259.6(4.3) $ & $ 55 $ & $ (22) \{21\} $ & $ 2 $ & $ (20) \{20\} $ & \cite{JYFLTRAP_GS} & $ 4 $ & $ (11) \{10\} $ & \cite{ISOL-Sn2} & $ -22 $ & $ (120) \{120\}^\textrm{a} $ & \cite{GSI_08} \\
$^{132}$Sn & 0.992\,568\,893(22) & $ -76\,549.0(2.8) $ & $ 5 $ & $ (14) \{14\} $ & $ -6 $ & $ (5) \{4\} $ & \cite{JYFLTRAP_GS} & $ -2 $ & $ (8) \{7\} $ & \cite{ISOL-Sn2} & $ $ & $  $ & \\
$^{133}$Sn & 1.000\,138\,949(29) & $ -70\,869.1(3.6) $ & $ 84 $ & $ (36) \{36\} $ & $ 5.3 $ & $ (4.3) \{2.4\} $ & \cite{JYFLTRAP_GS} & $ 22 $ & $ (23) \{23\} $ & \cite{ISOL-Sn2} & $ $ & $  $ & \\
$^{134}$Sn & 1.007\,698\,87(13) & $ -66\,444(16) $ & $ 350 $ & $ (100) \{100\} $ & $ 12 $ & $ (16) \{4\} $ & \cite{JYFLTRAP_GS} & $ -120 $ & $ (150) \{150\} $ & \cite{ISOL-Sn2} & $ $ & $  $ & \\
$^{135}$Sn & 1.015\,270\,38(28) & $ -60\,584(34) $ & $ 210 $ & $ (400) \{400\}\# $ & $ 48 $ & $ (35) \{3\} $ & \cite{JYFLTRAP_GS} & $ $ & $ $ & & $ $ & $  $ & \\
\noalign{\smallskip}                          
$^{131}$Sb & 0.985\,000\,799(84) & $ -81\,986(10) $ & $ 2 $ & $ (23) \{21\} $ & $ -3 $ & $ (10) \{2.1\} $ & \cite{JYFLTRAP_GS} & $ $ & $  $ & & $ $ & $  $ & \\
$^{132}$Sb$^\textrm{b}$ & 0.992\,543\,975(49) & $ -79\,633.8(6.1) $ & $ 40 $ & $ (16) \{14\} $ & $ 1.8 $ & $ (6.7) \{2.7\} $ & \cite{JYFLTRAP_GS} & $ $ & $  $ & & $ 236 $ & $ (124) \{124\}^\textrm{a} $ & \cite{GSI_08} \\
$^{133}$Sb$^\textrm{c}$ & 1.000\,073\,87(10) & $ -78\,921.3(7.6) $ & $ 21 $ & $ (27) \{25\} $ & $ 0 $ & $ (9) \{4\} $ & \cite{JYFLTRAP_GS} & $ $ & $  $ & & $ -22 $ & $ (24) \{23\} $ & \cite{GSI_12} \\
$^{134}$Sb$^\textrm{bd}$ & 1.007\,637\,742(82) & $ -74\,012(10) $ & $ 154 $ & $ (45) \{43\} $ & $ 10 $ & $ (10) \{2.1\} $ & \cite{JYFLTRAP_GS} & $ $ & $  $ & & $ $ & $  $ & \\
$^{135}$Sb & 1.015\,196\,795(53) & $ -69\,693.9(6.5) $ & $ 14 $ & $ (100) \{100\} $ & $ -4.3 $ & $ (7.1) \{2.9\} $ & \cite{JYFLTRAP_GS} & $ $ & $  $ & & $ 115 $ & $ (121) \{121\} $ & \cite{GSI_08} \\
$^{136}$Sb & 1.022\,763\,00(12) & $ -64\,491(15) $ & $ 390 $ & $ (300) \{300\}\# $ & $ 19 $ & $ (16) \{7\} $ & \cite{JYFLTRAP_GS} & $ $ & $  $ & & $ $ & $  $ & \\
$^{137}$Sb & 1.030\,322\,96(42) & $ -60\,061(52) $ & $ 200 $ & $ (400) \{400\}\# $ & $ $ & $  $ & & $ $ & $  $ & & $ $ & $  $ & \\
\noalign{\smallskip}                          
$^{133}$Te$^\textrm{b}$ & 1.000\,041\,770(52) & $ -82\,899.8(6.5) $ & $ 45 $ & $ (25) \{24\} $ & $ 38.4 $ & $ (6.8) \{2.2\} $ & \cite{JYFLTRAP_GS} & $ $ & $  $ & & $ $ & $  $ & \\
$^{135}$Te$^\textrm{c}$ & 1.015\,131\,888(18) & $ -77\,729.6(2.1) $ & $ 98 $ & $ (90) \{90\} $ & $ -1.7 $ & $ (3.3) \{2.6\} $ & \cite{JYFLTRAP_GS} & $ $ & $  $ & & $ -5 $ & $ (123) \{123\} $ & \cite{GSI_08} \\
$^{136}$Te$^\textrm{c}$ & 1.022\,682\,783(51) & $ -74\,423.3(3.7) $ & $ 2 $ & $ (45) \{45\} $ & $ 2.4 $ & $ (4.7) \{2.9\} $ & \cite{JYFLTRAP_GS} & $ $ & $  $ & & $ 45 $ & $ (23) \{23\} $ & \cite{GSI_12} \\
$^{137}$Te$^\textrm{c}$ & 1.030\,248\,309(31) & $ -69\,301.7(3.7) $ & $ 260 $ & $ (120) \{120\} $ & $ 2.5 $ & $ (4.5) \{2.5\} $ & \cite{JYFLTRAP_GS} & $ $ & $  $ & & $ -12 $ & $ (120) \{120\} $ & \cite{GSI_08} \\
$^{138}$Te & 1.037\,801\,624(61) & $ -65\,695.3(7.6) $ & $ 240 $ & $ (200) \{200\}\# $ & $ 1 $ & $ (9) \{5\} $ & \cite{JYFLTRAP_GS} & $ $ & $  $ & & $ 60 $ & $ (122) \{122\} $ & \cite{GSI_08} \\
$^{139}$Te & 1.045\,370\,26(13) & $ -60\,191(17) $ & $ 610 $ & $ (400) \{400\}\# $ & $ 14 $ & $ (17) \{4\} $ & \cite{JYFLTRAP_GS} & $ $ & $  $ & & $ $ & $  $ & \\
$^{140}$Te & 1.052\,923\,64(50) & $ -56\,577(62) $ & $ 380 $ & $ (300) \{300\}\# $ & $ -220 $ & $ (68) \{27\} $ & \cite{JYFLTRAP_GS} & $ $ & $  $ & & $ $ & $  $ & \\
\noalign{\smallskip}                          
$^{133}$I$^\textrm{e}$ & 1.000\,017\,873(52) & $ -85\,858.2(6.4) $ & $ 28 $ & $ (8) \{5\} $ & $ $ & $  $ & & $ $ & $  $ & & $ $ & $  $ & \\
$^{134}$I$^\textrm{b}$ & 1.007\,556\,731(51) & $ -84\,040.8(6.4) $ & $ 32 $ & $ (10) \{8\} $ & $ $ & $  $ & & $ $ & $  $ & & $ $ & $  $ & \\
$^{135}$I$^\textrm{c}$ & 1.015\,083\,028(17) & $ -83\,778.9(2.0) $ & $ 11 $ & $ (8) \{7\} $ & $ $ & $  $ & & $ $ & $  $ & & $ $ & $  $ & \\
$^{139}$I$^\textrm{c}$ & 1.045\,303\,381(32) & $ -68\,470.7(4.0) $ & $ 367 $ & $ (31) \{31\} $ & $ $ & $  $ & & $ $ & $  $ & & $ 56 $ & $ (121) \{121\} $ & \cite{GSI_08} \\
$^{140}$I & 1.052\,866\,85(10) & $ -63\,606(13) $ & $ 670 $ & $ (200) \{200\}\# $ & $ $ & $  $ & & $ $ & $  $ & & $ -10 $ & $ (122) \{121\} $ & \cite{GSI_08} \\
$^{141}$I & 1.060\,420\,75(13) & $ -59\,927(16) $ & $ 590 $ & $ (200) \{200\}\# $ & $ $ & $  $ & & $ $ & $  $ & & $ 374 $ & $ (130) \{129\} $ & \cite{GSI_08} \\
\noalign{\smallskip}                          
$^{142}$Cs$^\textrm{c}$ & 1.067\,859\,61(17) & $ -70\,506.9(9.3) $ & $ 8 $ & $ (14) \{11\} $ & $ $ & $  $ & & $ 14 $ & $ (17) \{15\}^\textrm{f} $ & \cite{99Am05} & $ $ & $  $ & \\
$^{143}$Cs & 1.075\,406\,507(64) & $ -67\,676.3(7.9) $ & $ -5 $ & $ (25) \{24\} $ & $ $ & $  $ & & $ $ & $  $ & & $ $ & $  $ & \\
$^{144}$Cs & 1.082\,966\,39(25) & $ -63\,256(31) $ & $ 14 $ & $ (41) \{26\} $ & $ $ & $  $ & & $ $ & $  $ & & $ $ & $  $ & \\
$^{145}$Cs & 1.090\,516\,40(13) & $ -60\,057(16) $ & $ 0 $ & $ (19) \{11\} $ & $ $ & $  $ & & $ -5 $ & $ (19) \{11\} $ & \cite{ISOL-Cs08} & $ $ & $  $ & \\
$^{146}$Cs & 1.098\,078\,820(69) & $ -55\,323.2(8.6) $ & $ 297 $ & $ (72) \{71\} $ & $ $ & $  $ & & $ $ & $  $ & & $ $ & $  $ & \\
\noalign{\smallskip}\hline\hline\noalign{\smallskip}                       
\multicolumn{14}{l}{ $^\textrm{a}$ An unknown mixture of the ground and isomer states}\\ 
\multicolumn{14}{l}{ $^\textrm{b}$ Resolved from isomer, ground state mass shown}\\ 
\multicolumn{14}{l}{ $^\textrm{c}$ Mass is the combined result for the CPT measurement in~\cite{jv12prc} and the new CARIBU data}\\ 
\multicolumn{14}{l}{ $^\textrm{d}$ The value in \cite{jv12prc} has been superseded by CARIBU data due to resolution of the two states present}\\
\multicolumn{14}{l}{ $^\textrm{e}$ Distant isomer was not targeted; observed state is expected to be ground state}\\ 
\multicolumn{14}{l}{ $^\textrm{f}$ Result has been adjusted by us due to a change in the calibration value after \cite{99Am05}. See \cite{jv12prc} for details.}\\ 
\multicolumn{14}{l}{ $\#$ Indicates an extrapolated mass value in the AME03}\\ 
\end{tabular*} 
\end{table*}

Table~\ref{tbl:masses} shows the masses measured by the CPT spectrometer.  The known half-lives of these nuclides ranges down to 280~ms.  For six of the nuclides---$^{130,131}$Sn, $^{132,134}$Sb, $^{133}$Te, and $^{134}$I---isomeric states were also clearly observed in the trap, which will be discussed in a future publication.  For two isotopes, $^{130,131}$In, isomeric states are expected to be present in the beam, but the CPT was unable either to resolve the states or to identify additional peaks, so no identification of the measured state or states has been made.  Durations of the measurement excitations ranged from 0.1~s to 7~s, depending on the lifetime of the nuclide and the proximity of isomers in cyclotron frequency.  Total mass uncertainty ranges from 2.1 to 62~keV/c$^2$, with a median uncertainty of 8.6~keV/c$^2$.

In comparison with the 2003 Atomic Mass Evaluation (AME03) values~~\cite{AME03_1,*AME03_2}, the previously seen trend of direct mass measurement values being higher than the AME03 far from stability continues (see examples in~\cite{jv12prc,fa11prc,fission_mass_review,Jyfl_Sr,Jyfl_Tc,ISOL-Xe09}).  This is likely due to systematic flaws in the $\beta$-endpoint measurements which previously dominated mass information in this region~\cite{Pandemonium}.  Also listed in Table~\ref{tbl:masses} are comparisons to other direct measurements.  The agreement for the 7 nuclides measured by both the CPT and ISOLTRAP~\cite{99Am05,ISOL-Cs08,ISOL-Sn1,ISOL-Sn2} is excellent.  The 19 common ground-state measurements with JYFLTRAP~\cite{JYFLTRAP_GS} are mostly in agreement, but with 2 masses---of $^{133}$Te and $^{140}$Te---disagreeing by 5.6 and 3.2~$\sigma$, respectively.  Measurements of these two isotopes were repeated in separate runs at the CPT, and gave consistent results.  The known isomer in $^{133}$Te is well-resolved, and no known isomers are expected to interfere with the ground state measurements of either $^{133}$Te or $^{140}$Te.  The cause of the disagreements remains unknown.  Finally, the overlap with the 9 ground states measured in the FRS-ESR ring~\cite{GSI_08,GSI_12} are in fair agreement, with two---$^{136}$Te and $^{141}$I---disagreeing by 2.0 and 2.9~$\sigma$, respectively.

\emph{Astrophysics impact.}---The bulk of the neutron captures in the $r$ process are thought to occur on timescales of order $1$~s in environments of temperature $T\gtrsim 1$~GK and neutron density $n_n \gtrsim 10^{20}$~cm$^{-3}$~\cite{Qian_2003}.  In that time, seed material from the nickel region must, through neutron capture and $\beta$ decay, be processed up through the heavy elements to explain the observed elemental abundances.  Because $\beta$ decay is the process that moves material to higher proton numbers, the $\beta$-decay lifetimes of the nuclei involved determine how quickly the heaviest elements can be reached.  The balance between neutron capture $(n,\gamma)$ and photodissociation $(\gamma,n)$ reactions determines the distribution of isotope populations within each element in a high-temperature $r$-process.  In hotter environments, increased $(\gamma,n)$ rates push this distribution closer to stability, and thus to nuclides with longer $\beta$-decay lifetimes.  At higher neutron densities, higher $(n,\gamma)$ rates have the opposite effect.

Due to the scarcity of direct mass measurements on the $r$-process path, theoretical nuclear mass models are used to provide the $S_n$ inputs needed for $(\gamma,n)$ rate calculations.  Unfortunately, even the most accurate mass models have RMS mass errors near 500~keV/c$^2$ for known nuclei~\cite{FRDM2012,HFB21,Lunney_review}, which, as is shown below, is insufficient for accurate $r$-process simulations.  Some of the nuclides with masses measured here---particularly isotopes of Sn and Sb---are on the $r$-process path for certain environmental conditions.  For those paths the measured Sn isotopes are among the nuclides with the longest $\beta$-decay lifetimes and are in a position to limit the progress of material to the heaviest elements.  Never before has the waiting time at this critical point on the $r$-process path been calculated from directly measured accurate masses for realistic conditions.

We performed therefore a series of focused $r$-process simulations to determine the waiting times at Sn and Sb for a span of conditions, using both this new mass information and, for comparison, three commonly used mass models: the Finite Range Droplet Model of 1995 (FRDM95)~\cite{FRDM95}, the Extended Thomas-Fermi plus Strutinksy Integral with enhanced Quenching model (ETFSI-Q)~\cite{ETFSIQ}, and the recent Hartree-Fock-Bogoliubov-21 model (HFB-21)~\cite{HFB21}.  The code used is one modified from that used earlier by this group in a similar $rp$-process calculation~\cite{cl07prc}.  Within a given simulation run, the temperature and neutron density of the $r$-process environment are fixed.  An initial population of nuclei is set on the low-$N$ side at $N=81$, and then the $(n,\gamma)$, $(\gamma,n)$, and $\beta$-decay processes are allowed to run over all appreciably populated isotopes of the simulated element.  The $(n,\gamma)$ and $(\gamma,n)$ rates come from the mass-model-based calculations of Rauscher and Thielemann~\cite{Rauscher_rates_2000}, which accept $S_n$ values as input parameters for the $(\gamma,n)$ rates.  Because isotopes beyond those measured by the CPT participate to some extent for all simulated environmental conditions, AME03 masses were used for lower-$N$ isotopes, and mass model $S_n$ values for  higher-$N$.  To isolate the effect of the differences between each model and the experimental masses, simulations were run with each of the mass models appended to the CPT data.  Separate simulations were run with each mass model used as inputs.  $\beta$-decay half-lives were taken from NuBase~\cite{NuBase} where available, and model-based half-lives~\cite{Moller_QRPA_lifes} were used for $^{138}$Sn, $^{138}$Sb, and beyond, scaled to match smoothly onto the known values. The simulations were run until half of the material had $\beta$-decayed, providing a measure of the waiting time via this ``effective'' half-life in each condition.  Simulations were run at 1.5~GK as a representative case, and were repeated over a span of neutron densities until the new masses were no longer relevant.

\begin{figure}
\centering
\includegraphics[width=0.5\textwidth]{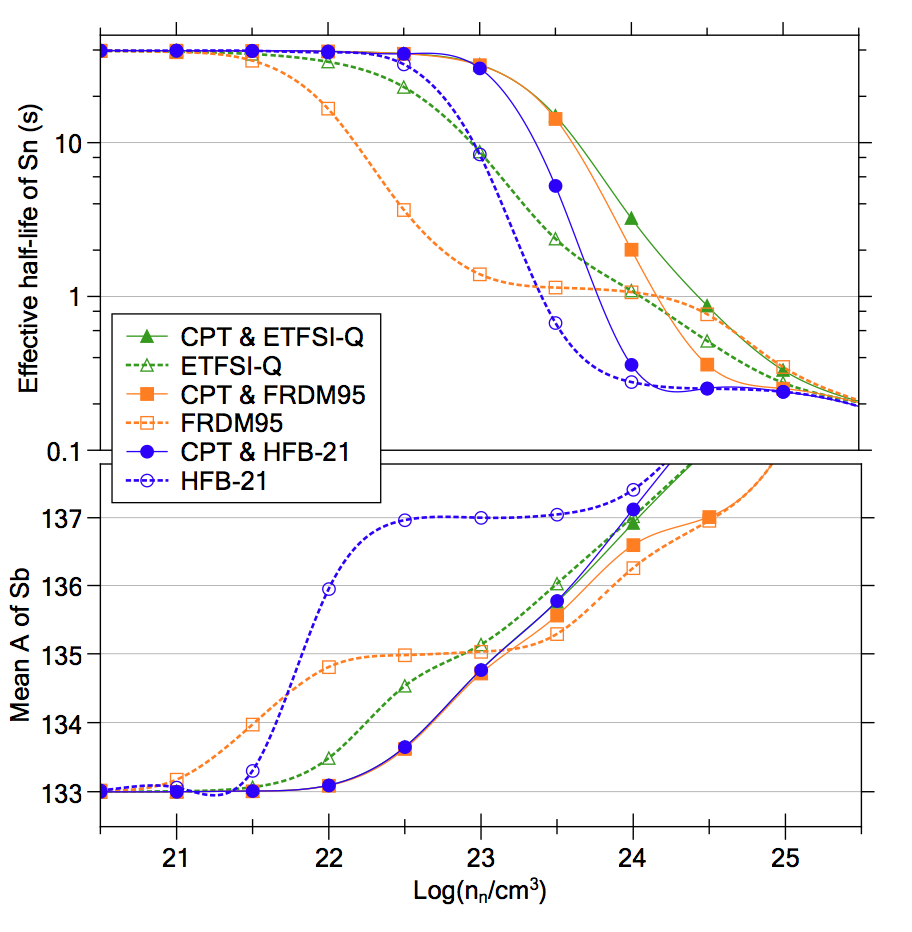}
\caption{\label{fig:sims}(Color online) Results of simulations performed at 1.5~GK. Top: The effective half-life of Sn versus neutron density.  Each mass model tested predicts a drop in time at lower density than in reality.  Bottom: The mean mass number of Sb versus neutron density.  The mass models all predict a breakaway from the neutron shell closure earlier than in reality, with HFB-21~\cite{HFB21} showing an unnaturally wide expanse in which $^{137}$Sb dominates.  See text for additional details.}
\end{figure}

Each mass-model-based simulation shows overly favorable prospects for moving past Sn, approaching a 1~s effective half-life at lower densities than the CPT-based simulations, as shown in the top of Fig.~\ref{fig:sims}.  This is due to the neutron separation energies of certain nuclei being too large in the mass models, impeding the $(\gamma,n)$ reactions that would push material to longer-lived nuclei.  In the ETFSI-Q case, this is caused mainly by a \hbox{350-keV} error in the $S_n$ of $^{133}$Sn.  The effect is more drastic for the FRDM95, which overestimates $S_n$ of both $^{133, 134}$Sn by 260 and 635~keV, respectively, followed by an underestimation for $^{135}$Sn by 340~keV. The combined effect of these is a prolonged span of densities for which $^{134}$Sn is the dominant isotope, with its \hbox{$1.1$-s} half-life setting the timescale.  The effect in HFB-21 starts at higher density than in the other models but is sharper.  The delay is due to a small undershot in the $S_n$ of $^{133}$Sn, and the rapid drop is started by a 604-keV error in $^{134}$Sn.  The mass models disagree with each other at $^{136}$Sn over a span of 839~keV, which induces the differences among the CPT-based simulations apparent in the plot. Simulations at other temperatures showed the same effects but at higher density for higher temperatures, as expected.

Similar effects emerged from the simulations of Sb, but the effective half-life is nearly always shorter at Sn than Sb over the relevant conditions.  Therefore plotted in the bottom of Fig.~\ref{fig:sims} is the mean mass number of the Sb populations at the end of each simulation, to illustrate the impact on $r$-process abundances.  There is a drastic effect in the HFB-21 simulations caused by an apparent lack of pairing around the neutron shell closure and consecutive 812 and 628~keV errors in the $S_n$ of $^{136,137}$Sb, which almost instantly drives all material from the $^{133}$Sb to $^{137}$Sb, an error of nearly 4 masses. Because this change is at low neutron densities the primary effect will be during freezeout, the time during which final $r$-process abundances (the only available $r$-process observable) are being determined.

These simulations provide a demonstration of the insufficiency of existing mass models for accurate $r$-process simulations. Progress in the development of mass models is slow, with the FRDM accuracy improving by $15\%$ over the last 17 years~\cite{FRDM2012} and the HFB models $32\%$ over 9 years~\cite{HFB1,HFB21}, for example.  Therefore without considerable improvements in nuclear theory, direct mass measurements may be the only sufficiently accurate source of $S_n$ inputs to $r$-process simulations for the foreseeable future.

In an actual $r$-process event, the temperature and density will both be evolving rapidly as the material expands and reactions take place.  These new results then provide a threshold of density which, for any given temperature, must be crossed in order to achieve a waiting time of $\sim 1$~s or less at Sn and move significant quantities of material to $Z>50$ before the $r$-process event ends.  Isotopes two neutrons farther from stability may be studied at CARIBU after installation of a 1 Ci fission source.  This would extend Sn measurements as far as $^{137}$Sn, which---with its \hbox{273-ms} half life---will set even stronger constraints on the $r$-process environment.

\emph{Acknowledgments.}---The authors acknowledge J. W. Truran and C. Ugalde for their helpful discussions regarding the simulations.  This work was performed under the auspices of NSERC, Canada, application number 216974, and the U.S. DOE, Office of Nuclear Physics, under Contract No. DE-AC02-06CH11357.

\bibliography{JVS,CPT,pastmass}

\begin{thebibliography}{47}%
\makeatletter
\providecommand \@ifxundefined [1]{%
 \@ifx{#1\undefined}
}%
\providecommand \@ifnum [1]{%
 \ifnum #1\expandafter \@firstoftwo
 \else \expandafter \@secondoftwo
 \fi
}%
\providecommand \@ifx [1]{%
 \ifx #1\expandafter \@firstoftwo
 \else \expandafter \@secondoftwo
 \fi
}%
\providecommand \natexlab [1]{#1}%
\providecommand \enquote  [1]{``#1''}%
\providecommand \bibnamefont  [1]{#1}%
\providecommand \bibfnamefont [1]{#1}%
\providecommand \citenamefont [1]{#1}%
\providecommand \href@noop [0]{\@secondoftwo}%
\providecommand \href [0]{\begingroup \@sanitize@url \@href}%
\providecommand \@href[1]{\@@startlink{#1}\@@href}%
\providecommand \@@href[1]{\endgroup#1\@@endlink}%
\providecommand \@sanitize@url [0]{\catcode `\\12\catcode `\$12\catcode
  `\&12\catcode `\#12\catcode `\^12\catcode `\_12\catcode `\%12\relax}%
\providecommand \@@startlink[1]{}%
\providecommand \@@endlink[0]{}%
\providecommand \url  [0]{\begingroup\@sanitize@url \@url }%
\providecommand \@url [1]{\endgroup\@href {#1}{\urlprefix }}%
\providecommand \urlprefix  [0]{URL }%
\providecommand \Eprint [0]{\href }%
\providecommand \doibase [0]{http://dx.doi.org/}%
\providecommand \selectlanguage [0]{\@gobble}%
\providecommand \bibinfo  [0]{\@secondoftwo}%
\providecommand \bibfield  [0]{\@secondoftwo}%
\providecommand \translation [1]{[#1]}%
\providecommand \BibitemOpen [0]{}%
\providecommand \bibitemStop [0]{}%
\providecommand \bibitemNoStop [0]{.\EOS\space}%
\providecommand \EOS [0]{\spacefactor3000\relax}%
\providecommand \BibitemShut  [1]{\csname bibitem#1\endcsname}%
\let\auto@bib@innerbib\@empty
\bibitem [{\citenamefont {{Van Duppen}}\ and\ \citenamefont
  {{Riisager}}(2011)}]{REX-ISOLDE}%
  \BibitemOpen
  \bibfield  {author} {\bibinfo {author} {\bibfnamefont {P.}~\bibnamefont {{Van
  Duppen}}} \ \bibnamefont {and}\ \bibinfo {author} {\bibfnamefont
  {K.}~\bibnamefont {{Riisager}}},\ }\href {\doibase
  10.1088/0954-3899/38/2/024005} {\bibfield  {journal} {\bibinfo  {journal} {J.
  Phys. G: Nucl. Part. Phys.}\ }\textbf {\bibinfo {volume} {38}},\ \bibinfo
  {pages} {024005} (\bibinfo {year} {2011})}\BibitemShut {NoStop}%
\bibitem [{\citenamefont {{{\"A}yst{\"o}}}(2001)}]{Aysto_IGISOL}%
  \BibitemOpen
  \bibfield  {author} {\bibinfo {author} {\bibfnamefont {J.}~\bibnamefont
  {{{\"A}yst{\"o}}}},\ }\href {\doibase 10.1016/S0375-9474(01)00923-X}
  {\bibfield  {journal} {\bibinfo  {journal} {Nucl. Phys. A}\ }\textbf
  {\bibinfo {volume} {693}},\ \bibinfo {pages} {477 } (\bibinfo {year}
  {2001})}\BibitemShut {NoStop}%
\bibitem [{\citenamefont {{Wada}}\ \emph {\textit{et~al.}}(2003)\citenamefont
  {{Wada}}, \citenamefont {{Ishida}}, \citenamefont {{Nakamura}}, \citenamefont
  {{Yamazaki}}, \citenamefont {{Kambara}}, \citenamefont {{Ohyama}},
  \citenamefont {{Kanai}}, \citenamefont {{Kojima}}, \citenamefont {{Nakai}},
  \citenamefont {{Ohshima}}, \citenamefont {{Yoshida}}, \citenamefont {{Kubo}},
  \citenamefont {{Matsuo}}, \citenamefont {{Fukuyama}}, \citenamefont
  {{Okada}}, \citenamefont {{Sonoda}}, \citenamefont {{Ohtani}}, \citenamefont
  {{Noda}}, \citenamefont {{Kawakami}},\ and\ \citenamefont
  {{Katayama}}}]{Wada_gas_catchers}%
  \BibitemOpen
  \bibfield  {author} {\bibinfo {author} {\bibfnamefont {M.}~\bibnamefont
  {{Wada}}} \bibnamefont {\textit{et~al.}},\ }\href {\doibase
  10.1016/S0168-583X(02)02151-1} {\bibfield  {journal} {\bibinfo  {journal}
  {Nucl. Instrum. Methods B}\ }\textbf {\bibinfo {volume} {204}},\ \bibinfo
  {pages} {570} (\bibinfo {year} {2003})}\BibitemShut {NoStop}%
\bibitem [{\citenamefont {{Savard}}\ \emph {\textit{et~al.}}(2003)\citenamefont
  {{Savard}}, \citenamefont {{Clark}}, \citenamefont {{Boudreau}},
  \citenamefont {{Buchinger}}, \citenamefont {{Crawford}}, \citenamefont
  {{Geissel}}, \citenamefont {{Greene}}, \citenamefont {{Gulick}},
  \citenamefont {{Heinz}}, \citenamefont {{Lee}}, \citenamefont {{Levand}},
  \citenamefont {{Maier}}, \citenamefont {{M{\" u}nzenberg}}, \citenamefont
  {{Scheidenberger}}, \citenamefont {{Seweryniak}}, \citenamefont {{Sharma}},
  \citenamefont {{Sprouse}}, \citenamefont {{Vaz}}, \citenamefont {{Wang}},
  \citenamefont {{Zabransky}}, \citenamefont {{Zhou}},\ and\ \citenamefont
  {{The S258 Collaboration}}}]{sa03nimb}%
  \BibitemOpen
  \bibfield  {author} {\bibinfo {author} {\bibfnamefont {G.}~\bibnamefont
  {{Savard}}} \bibnamefont {\textit{et~al.}},\ }\href@noop {} {\bibfield
  {journal} {\bibinfo  {journal} {Nucl. Instrum. Methods B}\ }\textbf {\bibinfo
  {volume} {204}},\ \bibinfo {pages} {582} (\bibinfo {year}
  {2003})}\BibitemShut {NoStop}%
\bibitem [{\citenamefont {{E. M. Burbidge, G. R. Burbidge, W. A. Fowler, and F.
  Hoyle}}(1957)}]{BBFH}%
  \BibitemOpen
  \bibfield  {author} {\bibinfo {author} {\bibnamefont {{E. M. Burbidge, G. R.
  Burbidge, W. A. Fowler, and F. Hoyle}}},\ }\href@noop {} {\bibfield
  {journal} {\bibinfo  {journal} {Rev. Mod. Phys.}\ }\textbf {\bibinfo {volume}
  {29}},\ \bibinfo {pages} {547} (\bibinfo {year} {1957})}\BibitemShut
  {NoStop}%
\bibitem [{\citenamefont {{Cowan}}\ \emph {\textit{et~al.}}(1991)\citenamefont
  {{Cowan}}, \citenamefont {{Thielemann}},\ and\ \citenamefont
  {{Truran}}}]{Cowan-rev}%
  \BibitemOpen
  \bibfield  {author} {\bibinfo {author} {\bibfnamefont {J.~J.}\ \bibnamefont
  {{Cowan}}}, \bibinfo {author} {\bibfnamefont {F.-K.}\ \bibnamefont
  {{Thielemann}}} \ \bibnamefont {and}\ \bibinfo {author} {\bibfnamefont
  {J.~W.}\ \bibnamefont {{Truran}}},\ }\href {\doibase
  10.1016/0370-1573(91)90070-3} {\bibfield  {journal} {\bibinfo  {journal}
  {Phys. Rep.}\ }\textbf {\bibinfo {volume} {208}},\ \bibinfo {pages} {267}
  (\bibinfo {year} {1991})}\BibitemShut {NoStop}%
\bibitem [{\citenamefont {{Qian}}(2003)}]{Qian_2003}%
  \BibitemOpen
  \bibfield  {author} {\bibinfo {author} {\bibfnamefont {Y.-Z.}\ \bibnamefont
  {{Qian}}},\ }\href {\doibase 10.1016/S0146-6410(02)00178-3} {\bibfield
  {journal} {\bibinfo  {journal} {Prog. Part. Nucl. Phys.}\ }\textbf {\bibinfo
  {volume} {50}},\ \bibinfo {pages} {153} (\bibinfo {year} {2003})}\BibitemShut
  {NoStop}%
\bibitem [{\citenamefont {{Arnould}}\ \emph
  {\textit{et~al.}}(2007)\citenamefont {{Arnould}}, \citenamefont {{Goriely}},\
  and\ \citenamefont {{Takahashi}}}]{Arnould_rev}%
  \BibitemOpen
  \bibfield  {author} {\bibinfo {author} {\bibfnamefont {M.}~\bibnamefont
  {{Arnould}}}, \bibinfo {author} {\bibfnamefont {S.}~\bibnamefont {{Goriely}}}
  \ \bibnamefont {and}\ \bibinfo {author} {\bibfnamefont {K.}~\bibnamefont
  {{Takahashi}}},\ }\href {\doibase 10.1016/j.physrep.2007.06.002} {\bibfield
  {journal} {\bibinfo  {journal} {Phys. Rep.}\ }\textbf {\bibinfo {volume}
  {450}},\ \bibinfo {pages} {97} (\bibinfo {year} {2007})}\BibitemShut
  {NoStop}%
\bibitem [{\citenamefont {{L. S. Brown and G. Gabrielse}}(1986)}]{Geonium}%
  \BibitemOpen
  \bibfield  {author} {\bibinfo {author} {\bibnamefont {{L. S. Brown and G.
  Gabrielse}}},\ }\href@noop {} {\bibfield  {journal} {\bibinfo  {journal}
  {Rev. Mod. Phys.}\ }\textbf {\bibinfo {volume} {58}},\ \bibinfo {pages} {233}
  (\bibinfo {year} {1986})}\BibitemShut {NoStop}%
\bibitem [{\citenamefont {{Fallis}}\ \emph {\textit{et~al.}}(2011)\citenamefont
  {{Fallis}}, \citenamefont {{Clark}}, \citenamefont {{Sharma}}, \citenamefont
  {{Savard}}, \citenamefont {{Buchinger}}, \citenamefont {{Caldwell}},
  \citenamefont {{Chaudhuri}}, \citenamefont {{Crawford}}, \citenamefont
  {{Deibel}}, \citenamefont {{Gulick}}, \citenamefont {{Hecht}}, \citenamefont
  {{Lascar}}, \citenamefont {{Lee}}, \citenamefont {{Levand}}, \citenamefont
  {{Li}}, \citenamefont {{Lundgren}}, \citenamefont {{Parikh}}, \citenamefont
  {{Russell}}, \citenamefont {{Scholte-van de Vorst}}, \citenamefont
  {{Scielzo}}, \citenamefont {{Segel}}, \citenamefont {{Sharma}}, \citenamefont
  {{Sinha}}, \citenamefont {{Sternberg}}, \citenamefont {{Sun}}, \citenamefont
  {{Tanihata}}, \citenamefont {{Van Schelt}}, \citenamefont {{Wang}},
  \citenamefont {{Wang}}, \citenamefont {{Wrede}},\ and\ \citenamefont
  {{Zhou}}}]{fa11prc}%
  \BibitemOpen
  \bibfield  {author} {\bibinfo {author} {\bibfnamefont {J.}~\bibnamefont
  {{Fallis}}} \bibnamefont {\textit{et~al.}},\ }\href {\doibase
  10.1103/PhysRevC.84.045807} {\bibfield  {journal} {\bibinfo  {journal} {Phys.
  Rev. C}\ }\textbf {\bibinfo {volume} {84}},\ \bibinfo {eid} {045807}
  (\bibinfo {year} {2011})}\BibitemShut {NoStop}%
\bibitem [{\citenamefont {Van~Schelt}(2012)}]{jv12phd}%
  \BibitemOpen
  \bibfield  {author} {\bibinfo {author} {\bibfnamefont {J.}~\bibnamefont
  {Van~Schelt}},\ }\emph {\bibinfo {title} {{Precision Mass Measurements of
  Neutron-Rich Nuclei, and Limitations on the \emph{\lowercase{r}}-Process
  Environment}}},\ \href@noop {} {Ph.D. thesis},\ \bibinfo  {school} {{The
  University of Chicago}} (\bibinfo {year} {2012}),\ \bibinfo {note}
  {{ProQuest/UMI 3526988}}\BibitemShut {NoStop}%
\bibitem [{\citenamefont {Lascar}(2012)}]{dl12phd}%
  \BibitemOpen
  \bibfield  {author} {\bibinfo {author} {\bibfnamefont {D.}~\bibnamefont
  {Lascar}},\ }\emph {\bibinfo {title} {{Precision Mass Measurements of Short
  Lived, Neutron Rich, R-Process Nuclei About the N=82 Waiting Point}}},\
  \href@noop {} {Ph.D. thesis},\ \bibinfo  {school} {{Northwestern University}}
  (\bibinfo {year} {2012}),\ \bibinfo {note} {{ProQuest/UMI
  3547913}}\BibitemShut {NoStop}%
\bibitem [{\citenamefont {Kankainen}\ \emph
  {\textit{et~al.}}(2012)\citenamefont {Kankainen}, \citenamefont {\"Ayst\"o},\
  and\ \citenamefont {Jokinen}}]{fission_mass_review}%
  \BibitemOpen
  \bibfield  {author} {\bibinfo {author} {\bibfnamefont {A.}~\bibnamefont
  {Kankainen}}, \bibinfo {author} {\bibfnamefont {J.}~\bibnamefont {\"Ayst\"o}}
  \ \bibnamefont {and}\ \bibinfo {author} {\bibfnamefont {A.}~\bibnamefont
  {Jokinen}},\ }\href {http://stacks.iop.org/0954-3899/39/i=9/a=093101}
  {\bibfield  {journal} {\bibinfo  {journal} {J. Phys. G: Nucl. Part. Phys.}\
  }\textbf {\bibinfo {volume} {39}},\ \bibinfo {pages} {093101} (\bibinfo
  {year} {2012})}\BibitemShut {NoStop}%
\bibitem [{\citenamefont {{Savard}}\ \emph {\textit{et~al.}}(2006)\citenamefont
  {{Savard}}, \citenamefont {{Wang}}, \citenamefont {{Sharma}}, \citenamefont
  {{Sharma}}, \citenamefont {{Clark}}, \citenamefont {{Boudreau}},
  \citenamefont {{Buchinger}}, \citenamefont {{Crawford}}, \citenamefont
  {{Greene}}, \citenamefont {{Gulick}}, \citenamefont {{Hecht}}, \citenamefont
  {{Lee}}, \citenamefont {{Levand}}, \citenamefont {{Scielzo}}, \citenamefont
  {{Trimble}}, \citenamefont {{Vaz}},\ and\ \citenamefont
  {{Zabransky}}}]{sa06ijmsip}%
  \BibitemOpen
  \bibfield  {author} {\bibinfo {author} {\bibfnamefont {G.}~\bibnamefont
  {{Savard}}} \bibnamefont {\textit{et~al.}},\ }\href {\doibase
  10.1016/j.ijms.2006.01.047} {\bibfield  {journal} {\bibinfo  {journal} {Int.
  J. Mass Spectrom.}\ }\textbf {\bibinfo {volume} {251}},\ \bibinfo {pages}
  {252} (\bibinfo {year} {2006})}\BibitemShut {NoStop}%
\bibitem [{\citenamefont {Van~Schelt}\ \emph
  {\textit{et~al.}}(2012)\citenamefont {Van~Schelt}, \citenamefont {Lascar},
  \citenamefont {Savard}, \citenamefont {Clark}, \citenamefont {Caldwell},
  \citenamefont {Chaudhuri}, \citenamefont {Fallis}, \citenamefont {Greene},
  \citenamefont {Levand}, \citenamefont {Li}, \citenamefont {Sharma},
  \citenamefont {Sternberg}, \citenamefont {Sun},\ and\ \citenamefont
  {Zabransky}}]{jv12prc}%
  \BibitemOpen
  \bibfield  {author} {\bibinfo {author} {\bibfnamefont {J.}~\bibnamefont
  {Van~Schelt}} \bibnamefont {\textit{et~al.}},\ }\href {\doibase
  10.1103/PhysRevC.85.045805} {\bibfield  {journal} {\bibinfo  {journal} {Phys.
  Rev. C}\ }\textbf {\bibinfo {volume} {85}},\ \bibinfo {pages} {045805}
  (\bibinfo {year} {2012})}\BibitemShut {NoStop}%
\bibitem [{\citenamefont {Savard}\ \emph {\textit{et~al.}}(2008)\citenamefont
  {Savard}, \citenamefont {Baker}, \citenamefont {Davids}, \citenamefont
  {Levand}, \citenamefont {Moore}, \citenamefont {Pardo}, \citenamefont
  {Vondrasek}, \citenamefont {Zabransky},\ and\ \citenamefont
  {Zinkann}}]{CARIBU}%
  \BibitemOpen
  \bibfield  {author} {\bibinfo {author} {\bibfnamefont {G.}~\bibnamefont
  {Savard}} \bibnamefont {\textit{et~al.}},\ }\href {\doibase
  10.1016/j.nimb.2008.05.091} {\bibfield  {journal} {\bibinfo  {journal} {Nucl.
  Instrum. Methods B}\ }\textbf {\bibinfo {volume} {266}},\ \bibinfo {pages}
  {4086 } (\bibinfo {year} {2008})}\BibitemShut {NoStop}%
\bibitem [{\citenamefont {{Davids}}\ and\ \citenamefont
  {{Peterson}}(2008)}]{CARIBU_IS}%
  \BibitemOpen
  \bibfield  {author} {\bibinfo {author} {\bibfnamefont {C.~N.}\ \bibnamefont
  {{Davids}}} \ \bibnamefont {and}\ \bibinfo {author} {\bibfnamefont
  {D.}~\bibnamefont {{Peterson}}},\ }\href {\doibase
  10.1016/j.nimb.2008.05.148} {\bibfield  {journal} {\bibinfo  {journal} {Nucl.
  Instrum. Methods B}\ }\textbf {\bibinfo {volume} {266}},\ \bibinfo {pages}
  {4449} (\bibinfo {year} {2008})}\BibitemShut {NoStop}%
\bibitem [{\citenamefont {{Gr{\"a}ff}}\ \emph
  {\textit{et~al.}}(1980)\citenamefont {{Gr{\"a}ff}}, \citenamefont
  {{Kalinowsky}},\ and\ \citenamefont {{Traut}}}]{Graff}%
  \BibitemOpen
  \bibfield  {author} {\bibinfo {author} {\bibfnamefont {G.}~\bibnamefont
  {{Gr{\"a}ff}}}, \bibinfo {author} {\bibfnamefont {H.}~\bibnamefont
  {{Kalinowsky}}} \ \bibnamefont {and}\ \bibinfo {author} {\bibfnamefont
  {J.}~\bibnamefont {{Traut}}},\ }\href {\doibase 10.1007/BF01414243}
  {\bibfield  {journal} {\bibinfo  {journal} {Z. Phys. A}\ }\textbf {\bibinfo
  {volume} {297}},\ \bibinfo {pages} {35} (\bibinfo {year} {1980})}\BibitemShut
  {NoStop}%
\bibitem [{\citenamefont {{Bollen}}\ \emph {\textit{et~al.}}(1990)\citenamefont
  {{Bollen}}, \citenamefont {{Moore}}, \citenamefont {{Savard}},\ and\
  \citenamefont {{Stolzenberg}}}]{Bollen_accuracy}%
  \BibitemOpen
  \bibfield  {author} {\bibinfo {author} {\bibfnamefont {G.}~\bibnamefont
  {{Bollen}}} \bibnamefont {\textit{et~al.}},\ }\href {\doibase
  10.1063/1.346185} {\bibfield  {journal} {\bibinfo  {journal} {J. Appl.
  Phys.}\ }\textbf {\bibinfo {volume} {68}},\ \bibinfo {pages} {4355} (\bibinfo
  {year} {1990})}\BibitemShut {NoStop}%
\bibitem [{\citenamefont {{M. K\"{o}nig, G. Bollen, H-J. Kluge, T. Otto, J.
  Szerypo}}(1995)}]{Konig_conversion}%
  \BibitemOpen
  \bibfield  {author} {\bibinfo {author} {\bibnamefont {{M. K\"{o}nig, G.
  Bollen, H-J. Kluge, T. Otto, J. Szerypo}}},\ }\href@noop {} {\bibfield
  {journal} {\bibinfo  {journal} {Int. J. Mass Spectrom. Ion Processes}\
  }\textbf {\bibinfo {volume} {142}},\ \bibinfo {pages} {95} (\bibinfo {year}
  {1995})}\BibitemShut {NoStop}%
\bibitem [{\citenamefont {George}\ \emph {\textit{et~al.}}(2011)\citenamefont
  {George}, \citenamefont {Blaum}, \citenamefont {Block}, \citenamefont
  {Breitenfeldt}, \citenamefont {Dworschak}, \citenamefont {Herfurth},
  \citenamefont {Herlert}, \citenamefont {Kowalska}, \citenamefont
  {Kretzschmar}, \citenamefont {Ramirez}, \citenamefont {Neidherr},
  \citenamefont {Schwarz},\ and\ \citenamefont {Schweikhard}}]{George_damping}%
  \BibitemOpen
  \bibfield  {author} {\bibinfo {author} {\bibfnamefont {S.}~\bibnamefont
  {George}} \bibnamefont {\textit{et~al.}},\ }\href {\doibase
  10.1016/j.ijms.2010.09.030} {\bibfield  {journal} {\bibinfo  {journal} {Int.
  J. Mass Spectrom.}\ }\textbf {\bibinfo {volume} {299}},\ \bibinfo {pages}
  {102} (\bibinfo {year} {2011})}\BibitemShut {NoStop}%
\bibitem [{\citenamefont {{Bollen}}\ \emph {\textit{et~al.}}(1992)\citenamefont
  {{Bollen}}, \citenamefont {{Kluge}}, \citenamefont {{K{\"o}nig}},
  \citenamefont {{Otto}}, \citenamefont {{Savard}}, \citenamefont
  {{Stolzenberg}}, \citenamefont {{Moore}}, \citenamefont {{Rouleau}},
  \citenamefont {{Audi}},\ and\ \citenamefont {{ISOLDE
  Collaboration}}}]{Bollen_isomer}%
  \BibitemOpen
  \bibfield  {author} {\bibinfo {author} {\bibfnamefont {G.}~\bibnamefont
  {{Bollen}}} \bibnamefont {\textit{et~al.}},\ }\href {\doibase
  10.1103/PhysRevC.46.R2140} {\bibfield  {journal} {\bibinfo  {journal} {\prc}\
  }\textbf {\bibinfo {volume} {46}},\ \bibinfo {pages} {R2140} (\bibinfo {year}
  {1992})}\BibitemShut {NoStop}%
\bibitem [{\citenamefont {{Wapstra}}\ \emph
  {\textit{et~al.}}(2003)\citenamefont {{Wapstra}}, \citenamefont {{Audi}},\
  and\ \citenamefont {{Thibault}}}]{AME03_1}%
  \BibitemOpen
  \bibfield  {author} {\bibinfo {author} {\bibfnamefont {A.~H.}\ \bibnamefont
  {{Wapstra}}}, \bibinfo {author} {\bibfnamefont {G.}~\bibnamefont {{Audi}}} \
  \bibnamefont {and}\ \bibinfo {author} {\bibfnamefont {C.}~\bibnamefont
  {{Thibault}}},\ }\href {\doibase 10.1016/j.nuclphysa.2003.11.002} {\bibfield
  {journal} {\bibinfo  {journal} {Nucl. Phys. A}\ }\textbf {\bibinfo {volume}
  {729}},\ \bibinfo {pages} {129} (\bibinfo {year} {2003})}\BibitemShut
  {NoStop}%
\bibitem [{\citenamefont {{Audi}}\ \emph
  {\textit{et~al.}}(2003{\natexlab{a}})\citenamefont {{Audi}}, \citenamefont
  {{Wapstra}},\ and\ \citenamefont {{Thibault}}}]{AME03_2}%
  \BibitemOpen
  \bibfield  {author} {\bibinfo {author} {\bibfnamefont {G.}~\bibnamefont
  {{Audi}}}, \bibinfo {author} {\bibfnamefont {A.~H.}\ \bibnamefont
  {{Wapstra}}} \ \bibnamefont {and}\ \bibinfo {author} {\bibfnamefont
  {C.}~\bibnamefont {{Thibault}}},\ }\href {\doibase
  10.1016/j.nuclphysa.2003.11.003} {\bibfield  {journal} {\bibinfo  {journal}
  {Nucl. Phys. A}\ }\textbf {\bibinfo {volume} {729}},\ \bibinfo {pages} {337}
  (\bibinfo {year} {2003}{\natexlab{a}})}\BibitemShut {NoStop}%
\bibitem [{\citenamefont {Hakala}\ \emph {\textit{et~al.}}(2012)\citenamefont
  {Hakala}, \citenamefont {Dobaczewski}, \citenamefont {Gorelov}, \citenamefont
  {Eronen}, \citenamefont {Jokinen}, \citenamefont {Kankainen}, \citenamefont
  {Kolhinen}, \citenamefont {Kortelainen}, \citenamefont {Moore}, \citenamefont
  {Penttil\"a}, \citenamefont {Rinta-Antila}, \citenamefont {Rissanen},
  \citenamefont {Saastamoinen}, \citenamefont {Sonnenschein},\ and\
  \citenamefont {\"Ayst\"o}}]{JYFLTRAP_GS}%
  \BibitemOpen
  \bibfield  {author} {\bibinfo {author} {\bibfnamefont {J.}~\bibnamefont
  {Hakala}} \bibnamefont {\textit{et~al.}},\ }\href {\doibase
  10.1103/PhysRevLett.109.032501} {\bibfield  {journal} {\bibinfo  {journal}
  {Phys. Rev. Lett.}\ }\textbf {\bibinfo {volume} {109}},\ \bibinfo {pages}
  {032501} (\bibinfo {year} {2012})}\BibitemShut {NoStop}%
\bibitem [{\citenamefont {{Ames}}\ \emph {\textit{et~al.}}(1999)\citenamefont
  {{Ames}}, \citenamefont {{Audi}}, \citenamefont {{Beck}}, \citenamefont
  {{Bollen}}, \citenamefont {{de Saint Simon}}, \citenamefont {{Jertz}},
  \citenamefont {{Kluge}}, \citenamefont {{Kohl}}, \citenamefont {{K{\"o}nig}},
  \citenamefont {{Lunney}}, \citenamefont {{Martel}}, \citenamefont {{Moore}},
  \citenamefont {{Otto}}, \citenamefont {{Patyk}}, \citenamefont
  {{Raimbault-Hartmann}}, \citenamefont {{Rouleau}}, \citenamefont {{Savard}},
  \citenamefont {{Schark}}, \citenamefont {{Schwarz}}, \citenamefont
  {{Schweikhard}}, \citenamefont {{Stolzenberg}}, \citenamefont {{Szerypo}},\
  and\ \citenamefont {{ISOLDE Collaboration}}}]{99Am05}%
  \BibitemOpen
  \bibfield  {author} {\bibinfo {author} {\bibfnamefont {F.}~\bibnamefont
  {{Ames}}} \bibnamefont {\textit{et~al.}},\ }\href {\doibase
  10.1016/S0375-9474(99)00111-6} {\bibfield  {journal} {\bibinfo  {journal}
  {Nucl. Phys. A}\ }\textbf {\bibinfo {volume} {651}},\ \bibinfo {pages} {3}
  (\bibinfo {year} {1999})}\BibitemShut {NoStop}%
\bibitem [{\citenamefont {Weber}\ \emph {\textit{et~al.}}(2008)\citenamefont
  {Weber}, \citenamefont {Audi}, \citenamefont {Beck}, \citenamefont {Blaum},
  \citenamefont {Bollen}, \citenamefont {Herfurth}, \citenamefont
  {Kellerbauer}, \citenamefont {Kluge}, \citenamefont {Lunney},\ and\
  \citenamefont {Schwarz}}]{ISOL-Cs08}%
  \BibitemOpen
  \bibfield  {author} {\bibinfo {author} {\bibfnamefont {C.}~\bibnamefont
  {Weber}} \bibnamefont {\textit{et~al.}},\ }\href {\doibase
  10.1016/j.nuclphysa.2007.12.014} {\bibfield  {journal} {\bibinfo  {journal}
  {Nucl. Phys. A}\ }\textbf {\bibinfo {volume} {803}},\ \bibinfo {pages} {1}
  (\bibinfo {year} {2008})}\BibitemShut {NoStop}%
\bibitem [{\citenamefont {Sikler}\ \emph {\textit{et~al.}}(2005)\citenamefont
  {Sikler}, \citenamefont {Audi}, \citenamefont {Beck}, \citenamefont {Blaum},
  \citenamefont {Bollen}, \citenamefont {Herfurth}, \citenamefont
  {Kellerbauer}, \citenamefont {Kluge}, \citenamefont {Lunney}, \citenamefont
  {Oinonen}, \citenamefont {Scheidenberger}, \citenamefont {Schwarz},\ and\
  \citenamefont {Szerypo}}]{ISOL-Sn1}%
  \BibitemOpen
  \bibfield  {author} {\bibinfo {author} {\bibfnamefont {G.}~\bibnamefont
  {Sikler}} \bibnamefont {\textit{et~al.}},\ }\href {\doibase
  10.1016/j.nuclphysa.2005.08.014} {\bibfield  {journal} {\bibinfo  {journal}
  {Nucl. Phys. A}\ }\textbf {\bibinfo {volume} {763}},\ \bibinfo {pages} {45}
  (\bibinfo {year} {2005})}\BibitemShut {NoStop}%
\bibitem [{\citenamefont {Dworschak}\ \emph
  {\textit{et~al.}}(2008)\citenamefont {Dworschak}, \citenamefont {Audi},
  \citenamefont {Blaum}, \citenamefont {Delahaye}, \citenamefont {George},
  \citenamefont {Hager}, \citenamefont {Herfurth}, \citenamefont {Herlert},
  \citenamefont {Kellerbauer}, \citenamefont {Kluge}, \citenamefont {Lunney},
  \citenamefont {Schweikhard},\ and\ \citenamefont {Yazidjian}}]{ISOL-Sn2}%
  \BibitemOpen
  \bibfield  {author} {\bibinfo {author} {\bibfnamefont {M.}~\bibnamefont
  {Dworschak}} \bibnamefont {\textit{et~al.}},\ }\href {\doibase
  10.1103/PhysRevLett.100.072501} {\bibfield  {journal} {\bibinfo  {journal}
  {Phys. Rev. Lett.}\ }\textbf {\bibinfo {volume} {100}},\ \bibinfo {pages}
  {072501} (\bibinfo {year} {2008})}\BibitemShut {NoStop}%
\bibitem [{\citenamefont {{Sun}}\ \emph {\textit{et~al.}}(2008)\citenamefont
  {{Sun}}, \citenamefont {{Kn{\"o}bel}}, \citenamefont {{Litvinov}},
  \citenamefont {{Geissel}}, \citenamefont {{Meng}}, \citenamefont {{Beckert}},
  \citenamefont {{Bosch}}, \citenamefont {{Boutin}}, \citenamefont {{Brandau}},
  \citenamefont {{Chen}}, \citenamefont {{Cullen}}, \citenamefont
  {{Dimopoulou}}, \citenamefont {{Fabian}}, \citenamefont {{Hausmann}},
  \citenamefont {{Kozhuharov}}, \citenamefont {{Litvinov}}, \citenamefont
  {{Mazzocco}}, \citenamefont {{Montes}}, \citenamefont {{M{\"u}nzenberg}},
  \citenamefont {{Musumarra}}, \citenamefont {{Nakajima}}, \citenamefont
  {{Nociforo}}, \citenamefont {{Nolden}}, \citenamefont {{Ohtsubo}},
  \citenamefont {{Ozawa}}, \citenamefont {{Patyk}}, \citenamefont {{Pla{\ss}}},
  \citenamefont {{Scheidenberger}}, \citenamefont {{Steck}}, \citenamefont
  {{Suzuki}}, \citenamefont {{Walker}}, \citenamefont {{Weick}}, \citenamefont
  {{Winckler}}, \citenamefont {{Winkler}},\ and\ \citenamefont
  {{Yamaguchi}}}]{GSI_08}%
  \BibitemOpen
  \bibfield  {author} {\bibinfo {author} {\bibfnamefont {B.}~\bibnamefont
  {{Sun}}} \bibnamefont {\textit{et~al.}},\ }\href {\doibase
  10.1016/j.nuclphysa.2008.08.013} {\bibfield  {journal} {\bibinfo  {journal}
  {Nucl. Phys. A}\ }\textbf {\bibinfo {volume} {812}},\ \bibinfo {pages} {1}
  (\bibinfo {year} {2008})}\BibitemShut {NoStop}%
\bibitem [{\citenamefont {Chen}\ \emph {\textit{et~al.}}(2012)\citenamefont
  {Chen}, \citenamefont {Plaß}, \citenamefont {Geissel}, \citenamefont
  {Knöbel}, \citenamefont {Kozhuharov}, \citenamefont {Litvinov}, \citenamefont
  {Patyk}, \citenamefont {Scheidenberger}, \citenamefont {Siegien-Iwaniuk},
  \citenamefont {Sun}, \citenamefont {Weick}, \citenamefont {Beckert},
  \citenamefont {Beller}, \citenamefont {Bosch}, \citenamefont {Boutin},
  \citenamefont {Caceres}, \citenamefont {Carroll}, \citenamefont {Cullen},
  \citenamefont {Cullen}, \citenamefont {Franzke}, \citenamefont {Gerl},
  \citenamefont {Górska}, \citenamefont {Jones}, \citenamefont {Kishada},
  \citenamefont {Kurcewicz}, \citenamefont {Litvinov}, \citenamefont {Liu},
  \citenamefont {Mandal}, \citenamefont {Montes}, \citenamefont {Münzenberg},
  \citenamefont {Nolden}, \citenamefont {Ohtsubo}, \citenamefont {Podolyák},
  \citenamefont {Propri}, \citenamefont {Rigby}, \citenamefont {Saito},
  \citenamefont {Saito}, \citenamefont {Shindo}, \citenamefont {Steck},
  \citenamefont {Walker}, \citenamefont {Williams}, \citenamefont {Winkler},
  \citenamefont {Wollersheim},\ and\ \citenamefont {Yamaguchi}}]{GSI_12}%
  \BibitemOpen
  \bibfield  {author} {\bibinfo {author} {\bibfnamefont {L.}~\bibnamefont
  {Chen}} \bibnamefont {\textit{et~al.}},\ }\href {\doibase
  10.1016/j.nuclphysa.2012.03.002} {\bibfield  {journal} {\bibinfo  {journal}
  {Nucl. Phys. A}\ }\textbf {\bibinfo {volume} {882}},\ \bibinfo {pages} {71 }
  (\bibinfo {year} {2012})}\BibitemShut {NoStop}%
\bibitem [{\citenamefont {Audi}\ \emph {\textit{et~al.}}(2012)\citenamefont
  {Audi}, \citenamefont {Wang}, \citenamefont {Wapstra}, \citenamefont
  {Kondev}, \citenamefont {MacCormick}, \citenamefont {Xu},\ and\ \citenamefont
  {Pfeiffer}}]{AME12_1}%
  \BibitemOpen
  \bibfield  {author} {\bibinfo {author} {\bibfnamefont {G.}~\bibnamefont
  {Audi}} \bibnamefont {\textit{et~al.}},\ }\href {\doibase
  10.1088/1674-1137/36/12/002} {\bibfield  {journal} {\bibinfo  {journal}
  {Chin. Phys.}\ }\textbf {\bibinfo {volume} {C36}},\ \bibinfo {pages} {1287}
  (\bibinfo {year} {2012})}\BibitemShut {NoStop}%
\bibitem [{\citenamefont {Wang}\ \emph {\textit{et~al.}}(2012)\citenamefont
  {Wang}, \citenamefont {Audi}, \citenamefont {Wapstra}, \citenamefont
  {Kondev}, \citenamefont {MacCormick}, \citenamefont {Xu},\ and\ \citenamefont
  {Pfeiffer}}]{AME12_2}%
  \BibitemOpen
  \bibfield  {author} {\bibinfo {author} {\bibfnamefont {M.}~\bibnamefont
  {Wang}} \bibnamefont {\textit{et~al.}},\ }\href {\doibase
  10.1088/1674-1137/36/12/003} {\bibfield  {journal} {\bibinfo  {journal}
  {Chin. Phys.}\ }\textbf {\bibinfo {volume} {C36}},\ \bibinfo {pages} {1603}
  (\bibinfo {year} {2012})}\BibitemShut {NoStop}%
\bibitem [{\citenamefont {{Hager}}\ \emph {\textit{et~al.}}(2006)\citenamefont
  {{Hager}}, \citenamefont {{Eronen}}, \citenamefont {{Hakala}}, \citenamefont
  {{Jokinen}}, \citenamefont {{Kolhinen}}, \citenamefont {{Kopecky}},
  \citenamefont {{Moore}}, \citenamefont {{Nieminen}}, \citenamefont
  {{Oinonen}}, \citenamefont {{Rinta-Antila}}, \citenamefont {{Szerypo}},\ and\
  \citenamefont {{{\"A}yst{\"o}}}}]{Jyfl_Sr}%
  \BibitemOpen
  \bibfield  {author} {\bibinfo {author} {\bibfnamefont {U.}~\bibnamefont
  {{Hager}}} \bibnamefont {\textit{et~al.}},\ }\href {\doibase
  10.1103/PhysRevLett.96.042504} {\bibfield  {journal} {\bibinfo  {journal}
  {\prl}\ }\textbf {\bibinfo {volume} {96}},\ \bibinfo {eid} {042504} (\bibinfo
  {year} {2006})}\BibitemShut {NoStop}%
\bibitem [{\citenamefont {{Hager}}\ \emph {\textit{et~al.}}(2007)\citenamefont
  {{Hager}}, \citenamefont {{Elomaa}}, \citenamefont {{Eronen}}, \citenamefont
  {{Hakala}}, \citenamefont {{Jokinen}}, \citenamefont {{Kankainen}},
  \citenamefont {{Rahaman}}, \citenamefont {{Rinta-Antila}}, \citenamefont
  {{Saastamoinen}}, \citenamefont {{Sonoda}},\ and\ \citenamefont
  {{{\"A}yst{\"o}}}}]{Jyfl_Tc}%
  \BibitemOpen
  \bibfield  {author} {\bibinfo {author} {\bibfnamefont {U.}~\bibnamefont
  {{Hager}}} \bibnamefont {\textit{et~al.}},\ }\href {\doibase
  10.1103/PhysRevC.75.064302} {\bibfield  {journal} {\bibinfo  {journal}
  {\prc}\ }\textbf {\bibinfo {volume} {75}},\ \bibinfo {eid} {064302} (\bibinfo
  {year} {2007})}\BibitemShut {NoStop}%
\bibitem [{\citenamefont {{Neidherr}}\ \emph
  {\textit{et~al.}}(2009)\citenamefont {{Neidherr}}, \citenamefont {{Cakirli}},
  \citenamefont {{Audi}}, \citenamefont {{Beck}}, \citenamefont {{Blaum}},
  \citenamefont {{B{\"o}hm}}, \citenamefont {{Breitenfeldt}}, \citenamefont
  {{Casten}}, \citenamefont {{George}}, \citenamefont {{Herfurth}},
  \citenamefont {{Herlert}}, \citenamefont {{Kellerbauer}}, \citenamefont
  {{Kowalska}}, \citenamefont {{Lunney}}, \citenamefont {{Minaya-Ramirez}},
  \citenamefont {{Naimi}}, \citenamefont {{Rosenbusch}}, \citenamefont
  {{Schwarz}},\ and\ \citenamefont {{Schweikhard}}}]{ISOL-Xe09}%
  \BibitemOpen
  \bibfield  {author} {\bibinfo {author} {\bibfnamefont {D.}~\bibnamefont
  {{Neidherr}}} \bibnamefont {\textit{et~al.}},\ }\href {\doibase
  10.1103/PhysRevC.80.044323} {\bibfield  {journal} {\bibinfo  {journal}
  {\prc}\ }\textbf {\bibinfo {volume} {80}},\ \bibinfo {eid} {044323} (\bibinfo
  {year} {2009})}\BibitemShut {NoStop}%
\bibitem [{\citenamefont {Hardy}\ \emph {\textit{et~al.}}(1977)\citenamefont
  {Hardy}, \citenamefont {Carraz}, \citenamefont {Jonson},\ and\ \citenamefont
  {Hansen}}]{Pandemonium}%
  \BibitemOpen
  \bibfield  {author} {\bibinfo {author} {\bibfnamefont {J.~C.}\ \bibnamefont
  {Hardy}} \bibnamefont {\textit{et~al.}},\ }\href {\doibase
  10.1016/0370-2693(77)90223-4} {\bibfield  {journal} {\bibinfo  {journal}
  {Phys. Lett. B}\ }\textbf {\bibinfo {volume} {71}},\ \bibinfo {pages} {307 }
  (\bibinfo {year} {1977})}\BibitemShut {NoStop}%
\bibitem [{\citenamefont {M\"oller}\ \emph {\textit{et~al.}}(2012)\citenamefont
  {M\"oller}, \citenamefont {Myers}, \citenamefont {Sagawa},\ and\
  \citenamefont {Yoshida}}]{FRDM2012}%
  \BibitemOpen
  \bibfield  {author} {\bibinfo {author} {\bibfnamefont {P.}~\bibnamefont
  {M\"oller}} \bibnamefont {\textit{et~al.}},\ }\href {\doibase
  10.1103/PhysRevLett.108.052501} {\bibfield  {journal} {\bibinfo  {journal}
  {Phys. Rev. Lett.}\ }\textbf {\bibinfo {volume} {108}},\ \bibinfo {pages}
  {052501} (\bibinfo {year} {2012})}\BibitemShut {NoStop}%
\bibitem [{\citenamefont {Goriely}\ \emph {\textit{et~al.}}(2010)\citenamefont
  {Goriely}, \citenamefont {Chamel},\ and\ \citenamefont {Pearson}}]{HFB21}%
  \BibitemOpen
  \bibfield  {author} {\bibinfo {author} {\bibfnamefont {S.}~\bibnamefont
  {Goriely}}, \bibinfo {author} {\bibfnamefont {N.}~\bibnamefont {Chamel}} \
  \bibnamefont {and}\ \bibinfo {author} {\bibfnamefont {J.~M.}\ \bibnamefont
  {Pearson}},\ }\href {\doibase 10.1103/PhysRevC.82.035804} {\bibfield
  {journal} {\bibinfo  {journal} {Phys. Rev. C}\ }\textbf {\bibinfo {volume}
  {82}},\ \bibinfo {pages} {035804} (\bibinfo {year} {2010})}\BibitemShut
  {NoStop}%
\bibitem [{\citenamefont {Lunney}\ \emph {\textit{et~al.}}(2003)\citenamefont
  {Lunney}, \citenamefont {Pearson},\ and\ \citenamefont
  {Thibault}}]{Lunney_review}%
  \BibitemOpen
  \bibfield  {author} {\bibinfo {author} {\bibfnamefont {D.}~\bibnamefont
  {Lunney}}, \bibinfo {author} {\bibfnamefont {J.~M.}\ \bibnamefont {Pearson}}
  \ \bibnamefont {and}\ \bibinfo {author} {\bibfnamefont {C.}~\bibnamefont
  {Thibault}},\ }\href {\doibase 10.1103/RevModPhys.75.1021} {\bibfield
  {journal} {\bibinfo  {journal} {Rev. Mod. Phys.}\ }\textbf {\bibinfo {volume}
  {75}},\ \bibinfo {pages} {1021} (\bibinfo {year} {2003})}\BibitemShut
  {NoStop}%
\bibitem [{\citenamefont {{M{\"o}ller}}\ \emph
  {\textit{et~al.}}(1995)\citenamefont {{M{\"o}ller}}, \citenamefont {{Nix}},
  \citenamefont {{Myers}},\ and\ \citenamefont {{Swiatecki}}}]{FRDM95}%
  \BibitemOpen
  \bibfield  {author} {\bibinfo {author} {\bibfnamefont {P.}~\bibnamefont
  {{M{\"o}ller}}} \bibnamefont {\textit{et~al.}},\ }\href {\doibase
  10.1006/adnd.1995.1002} {\bibfield  {journal} {\bibinfo  {journal} {At. Data
  Nucl. Data Tables}\ }\textbf {\bibinfo {volume} {59}},\ \bibinfo {pages}
  {185} (\bibinfo {year} {1995})}\BibitemShut {NoStop}%
\bibitem [{\citenamefont {{Pearson}}\ \emph
  {\textit{et~al.}}(1996)\citenamefont {{Pearson}}, \citenamefont {{Nayak}},\
  and\ \citenamefont {{Goriely}}}]{ETFSIQ}%
  \BibitemOpen
  \bibfield  {author} {\bibinfo {author} {\bibfnamefont {J.~M.}\ \bibnamefont
  {{Pearson}}}, \bibinfo {author} {\bibfnamefont {R.~C.}\ \bibnamefont
  {{Nayak}}} \ \bibnamefont {and}\ \bibinfo {author} {\bibfnamefont
  {S.}~\bibnamefont {{Goriely}}},\ }\href {\doibase
  10.1016/0370-2693(96)01071-4} {\bibfield  {journal} {\bibinfo  {journal}
  {Phys. Lett. B}\ }\textbf {\bibinfo {volume} {387}},\ \bibinfo {pages} {455}
  (\bibinfo {year} {1996})}\BibitemShut {NoStop}%
\bibitem [{\citenamefont {{Clark}}\ \emph {\textit{et~al.}}(2007)\citenamefont
  {{Clark}}, \citenamefont {{Sharma}}, \citenamefont {{Savard}}, \citenamefont
  {{Levand}}, \citenamefont {{Wang}}, \citenamefont {{Zhou}}, \citenamefont
  {{Blank}}, \citenamefont {{Buchinger}}, \citenamefont {{Crawford}},
  \citenamefont {{Gulick}}, \citenamefont {{Lee}}, \citenamefont
  {{Seweryniak}},\ and\ \citenamefont {{Trimble}}}]{cl07prc}%
  \BibitemOpen
  \bibfield  {author} {\bibinfo {author} {\bibfnamefont {J.~A.}\ \bibnamefont
  {{Clark}}} \bibnamefont {\textit{et~al.}},\ }\href {\doibase
  10.1103/PhysRevC.75.032801} {\bibfield  {journal} {\bibinfo  {journal} {Phys.
  Rev. C}\ }\textbf {\bibinfo {volume} {75}},\ \bibinfo {eid} {032801}
  (\bibinfo {year} {2007})}\BibitemShut {NoStop}%
\bibitem [{\citenamefont {{Rauscher}}\ and\ \citenamefont
  {{Thielemann}}(2000)}]{Rauscher_rates_2000}%
  \BibitemOpen
  \bibfield  {author} {\bibinfo {author} {\bibfnamefont {T.}~\bibnamefont
  {{Rauscher}}} \ \bibnamefont {and}\ \bibinfo {author} {\bibfnamefont {F.-K.}\
  \bibnamefont {{Thielemann}}},\ }\href {\doibase 10.1006/adnd.2000.0834}
  {\bibfield  {journal} {\bibinfo  {journal} {At. Data Nucl. Data Tables}\
  }\textbf {\bibinfo {volume} {75}},\ \bibinfo {pages} {1} (\bibinfo {year}
  {2000})}\BibitemShut {NoStop}%
\bibitem [{\citenamefont {{Audi}}\ \emph
  {\textit{et~al.}}(2003{\natexlab{b}})\citenamefont {{Audi}}, \citenamefont
  {{Bersillon}}, \citenamefont {{Blachot}},\ and\ \citenamefont
  {{Wapstra}}}]{NuBase}%
  \BibitemOpen
  \bibfield  {author} {\bibinfo {author} {\bibfnamefont {G.}~\bibnamefont
  {{Audi}}} \bibnamefont {\textit{et~al.}},\ }\href {\doibase
  10.1016/j.nuclphysa.2003.11.001} {\bibfield  {journal} {\bibinfo  {journal}
  {Nucl. Phys. A}\ }\textbf {\bibinfo {volume} {729}},\ \bibinfo {pages} {3}
  (\bibinfo {year} {2003}{\natexlab{b}})}\BibitemShut {NoStop}%
\bibitem [{\citenamefont {{M{\"o}ller}}\ \emph
  {\textit{et~al.}}(2003)\citenamefont {{M{\"o}ller}}, \citenamefont
  {{Pfeiffer}},\ and\ \citenamefont {{Kratz}}}]{Moller_QRPA_lifes}%
  \BibitemOpen
  \bibfield  {author} {\bibinfo {author} {\bibfnamefont {P.}~\bibnamefont
  {{M{\"o}ller}}}, \bibinfo {author} {\bibfnamefont {B.}~\bibnamefont
  {{Pfeiffer}}} \ \bibnamefont {and}\ \bibinfo {author} {\bibfnamefont {K.-L.}\
  \bibnamefont {{Kratz}}},\ }\href {\doibase 10.1103/PhysRevC.67.055802}
  {\bibfield  {journal} {\bibinfo  {journal} {Phys. Rev. C}\ }\textbf {\bibinfo
  {volume} {67}},\ \bibinfo {eid} {055802} (\bibinfo {year} {2003})},\ \bibinfo
  {note} {values retrieved from
  http://t16web.lanl.gov/Moller/publications/tpnff.dat}\BibitemShut {NoStop}%
\bibitem [{\citenamefont {Samyn}\ \emph {\textit{et~al.}}(2002)\citenamefont
  {Samyn}, \citenamefont {Goriely}, \citenamefont {Heenen}, \citenamefont
  {Pearson},\ and\ \citenamefont {Tondeur}}]{HFB1}%
  \BibitemOpen
  \bibfield  {author} {\bibinfo {author} {\bibfnamefont {M.}~\bibnamefont
  {Samyn}} \bibnamefont {\textit{et~al.}},\ }\href {\doibase
  10.1016/S0375-9474(01)01316-1} {\bibfield  {journal} {\bibinfo  {journal}
  {Nucl. Phys. A}\ }\textbf {\bibinfo {volume} {700}},\ \bibinfo {pages} {142 }
  (\bibinfo {year} {2002})}\BibitemShut {NoStop}%
\end{thebibliography}%

\end{document}